\documentstyle[12pt]{article}
\def \lesssim
{\,\lower 3 pt \hbox{\vbox{\hbox{$<$} \kern -9 pt \hbox{$\sim$}}}\,}
\def \gtrsim
{\,\lower 3 pt \hbox{\vbox{\hbox{$>$} \kern -9 pt \hbox{$\sim$}}}\,}
\input epsf.tex
\def\DESepsf(#1 width #2){\epsfxsize=#2 \epsfbox{#1}}
\begin{document}

\title{Factorization is not violated}

\author{John C.\ Collins
    \\{\it Department of Physics, Penn State University
        104 Davey Lab.}
    \\{\it University Park, PA 16802, U.S.A.}
\and
    Davison E.\ Soper
    \\{\it Institute of Theoretical Science,
        University of Oregon}
    \\{\it Eugene, OR 97403, U.S.A.}
\and
    George Sterman
    \\{\it Institute for Theoretical Physics,
       State University of New York}
    \\{\it  Stony Brook, NY 11794-3840, U.S.A.}
}

\date{2 June 1998}
\maketitle

\begin{abstract}
We show that existing proofs of factorization imply the cancellation of
certain multiladder contributions that Gotsman, Levin, and Maor had
suggested would invalidate the basic factorization theorem in QCD.  No
modifications of the original argument are necessary, although the
details of the example offer useful insight into the mechanisms of
factorization.
\end{abstract}

\section{Introduction}

Factorization theorems are at the heart of the theory of high-energy
inclusive processes involving a large momentum transfer. Examples
include the inclusive production in hadron-hadron collisions of new,
heavy particles, such as the Higgs boson or the supersymmetric partners
of the currently known particles.  These theorems apply to {\em
inclusive} cross sections for the production of a set $F$ of heavy
particles or of a system of jets with a total mass $Q$,
\begin{equation}
A+B\rightarrow F(Q)+X\, .
\label{ABFX}
\end{equation}
The theorem assumes that $F$ is defined in such a way that all final
states that differ by the emission of soft hadrons and by the collinear
rearrangement of momenta between outgoing light hadrons are summed
over. It is important here that $Q$ be large. One usually assumes that
$Q$ is of the order of the $AB$ invariant mass $\sqrt s$. For $1\ {\rm
GeV} \ll Q \ll \sqrt s$, one should sum logarithms of $\sqrt s/Q$ to
improve the usefulness of the perturbative expansion used in the
calculations.

According to the factorization theorem, the cross section for
(\ref{ABFX}) may be written as a convolution of parton distributions
and perturbative hard scattering functions,
\begin{equation}
\sigma _{AB}
=
\sum _{ab} \int dx_{a}dx_{b}\,
\phi _{a/A}(x_{a},\mu ^{2})\,
\phi _{b/B}(x_{b},\mu ^{2})\,
\hat{\sigma }_{ab}\!\left({Q^{2}\over x_{a}x_{b}s},{Q\over \mu} ,\alpha
_{s}(\mu )\right)
\, \left(1 + {\cal O}\left({1\over Q^{P}}\right) \right) \, .
\label{factth}
\end{equation}
Here, the nonperturbative functions $\phi _{c/C}(x_{c},\mu ^{2})$ give
the distribution of parton $c$ in hadron $C$ as a function of the
longitudinal momentum fraction $x_{c}$, evolved to factorization scale
$\mu $. The hard-scattering function $\hat{\sigma }_{ab}$  is computed
in perturbation theory up to some order, $\alpha_s^N$, leaving
corrections of order $\alpha_s^{N+1}$. As indicated, corrections to the
factorization formula are suppressed by a positive power, $P$, of the
large mass.

Despite their central role, the foundations of the factorization
theorems have received rather little theoretical attention in recent
years beyond the arguments given in Ref.~\cite{CSS85,CSS88} and reviewed
in Ref.~\cite{CSSrv}.  An exception is a recent paper by Gotsman, Levin,
and Maor \cite{levinetal}, which discusses the possibility of corrections
to the right-hand side of Eq.~(\ref{factth}) that are not
power-suppressed at all. These conjectured corrections are associated
with the exchange of ladders of the reggeon type. It is our aim in this
paper to show that such exchanges do {\it not} require a revision of
Eq.~(\ref{factth}), and that, indeed, in an inclusive cross section of
the type described above, contributions from the regions of momentum
space identified in \cite{levinetal} cancel.  We shall rely heavily on
the discussion of Ref.~\cite{CSS88}, and will not find it necessary to
modify the reasoning presented there.  Indeed, the discussion of that
paper already includes the treatment of the effect identified in
Ref.~\cite{levinetal}, to all orders in perturbation theory.
Nevertheless, since Ref.~\cite{CSS88} does not include low-order
examples, the treatment of a rapidity-ordered ladder will, we hope,
illustrate and perhaps clarify the proof.

\section{Ladder Diagrams in Inclusive Cross Sections}

We consider the cross section $d\sigma /dy$ for the production of a
Higgs boson with rapidity $y$ in the collision of two hadrons with c.m.\
energy $\sqrt s$. Our hadrons each consist of a simple quark-antiquark
pair with a pointlike coupling to the hadron field. The hard scale in
the problem is the Higgs boson mass, which we denote by $Q$. We may
have $s \sim Q^{2}$ or $s \gg Q^{2}$. Our analysis works in either
case.

\begin{figure}
\centerline{ \DESepsf(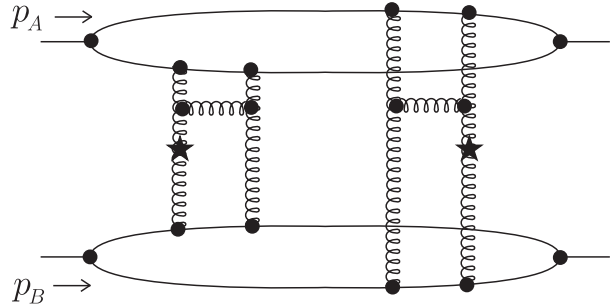 width 8 cm) }
\caption{ The uncut version of the graph analyzed in this paper. }
\label{basic}
\end{figure}

We analyze cut Feynman diagrams for Higgs boson production that arise
from the uncut graph shown in Fig.~\ref{basic}. This choice of graph is
motivated by the suggestion of Gotsman {\it et al.} \cite{levinetal},
who examine graphs similar to this. This graph contains enough of the
physics to be instructive, while it is of low enough order to be
treated in detail. 
We hope that it will be evident that the argument given below
for the graph of Fig.~\ref{basic} could be applied to more complicated
graphs.

\begin{figure}
\centerline{ \DESepsf(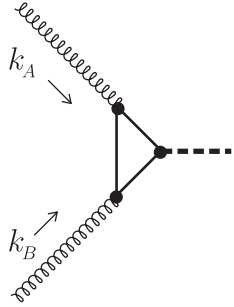 width 3 cm) }
\caption{ The hard subgraph, represented by a star in our graphs. }
\label{hard}
\end{figure}

In  Fig.~\ref{basic}, the star represents the hard scattering matrix
element for ${\rm gluon} + {\rm gluon} \to {\rm Higgs\ boson}$. The
lowest order diagram for this, containing a top quark loop, is shown in
Fig.~\ref{hard}.
We shall treat only the case where Higgs production appears
on the ``outer" sides of the ladders, as in Fig.\ \ref{basic}.  Moving
the starred vertex to the other side of either or both ladders
\cite{levinetal} leads
to diagrams that can be treated by exactly the same methods as Fig.\
\ref{basic}.

We take $p^{\mu }_{A} = \left(p^{+} _{A}, 0, {\bf 0} \right)$ and
$p^{\mu }_{B} = \left(0,p^{-} _{B}, {\bf 0} \right)$, where $p^\pm =
(p^0 \pm p^3)/\sqrt 2$. Then $2 p_{A}^{+} p_{B}^{-} = s$.  Let the Higgs
boson momentum be $q^{\mu }$, so $q_{\mu }q^{\mu }= Q^{2}$. It is
helpful to have a specific reference frame in mind. Let us choose the
reference frame in which $q^{+} = q^{-}$.

There are various integration regions that make leading contributions
to the cross section. We consider the region in which all transverse
momenta are of order $m$ ($m \sim 1 \ {\rm GeV} \ll Q$) and in which
all lines outside the hard scattering have virtuality at most of order
$m^{2}$. We take quark masses to be of order $m$.  The gluons entering
the hard interaction have momenta $k^{\mu }_{A} = \left(k^{+}_{A},
k^{-}_{A}, {\bf k}_{A} \right)$, $k^{\mu }_{B} = \left(k^{+}_{B},
k^{-}_{B}, {\bf k}_{B} \right)$ where $k_A^+$ and $k_B^-$ are of order
$Q$. The transverse components ${\bf k}_{A}^{\mu}$ and ${\bf k}_{B}^{\mu
}$ are of order $m$. Then $k_{A}^{-}$ and $k_{B}^{+}$ are much smaller
than $m$, of order $m^2/Q$ in order that $k_A^2 \sim k_B^2 \sim m^2$. In
the integration region of interest, all of the lines in the upper part
of the graph carry large $+$ momenta, with positive $+$ momentum carried
from left to right through the graph. The transverse momentum components
are all of order $m$ and the $-$ components are all small. Now consider
the bottom part of the graph, not including the two gluons exchanged
between top and bottom. All of the of the lines here carry large $-$
momenta, with positive $-$ momentum carried from left to right through
the graph. The transverse momentum components are all of order $m$
and the $+$ components are all small. We shall refer to this as the
``ladder region" of momentum space.

\begin{figure}
\centerline{ \DESepsf(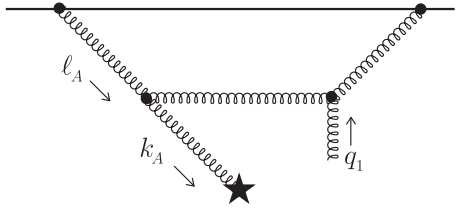 width 6 cm) }
\caption{ Labeling of momenta for one of the ladder subdiagrams. }
\label{nascent}
\end{figure}

There are two one-rung ladders in the diagram. For the left hand ladder,
for instance, we choose the labeling shown in Fig.~\ref{nascent}. As
mentioned above, we work in the integration region with $\ell _{A}^{T}
\sim k_{A}^{T} \sim q_{1}^{T} \sim m$.  We take $k^{+}_{A} = x\,p^{+}_{A}$
and $\ell ^{+}_{A} = z\,p^{+}_{A}$ with $x_{A} <z <1$. If $x_{A}
\ll 1$ then $z$-integration over this range can give a large logarithm
$\ln(1/x_{A})$, where the logarithm comes from the strongly ordered region
$x_{A} \ll z \ll 1$.  However, we do not need strong ordering. What is
important for us is that $\ell _{A}^{+}$, $k_{A}^{+}$ and $\ell _{A}^{+}
- k_{A}^{+}$ are large and positive, at least of order $Q$.

By assumption, we are in a region of momentum where the exchanged gluon
with momentum $q_{1}^{\mu }$ has $q_{1}^{T}\sim m$ and all the lines
outside the hard scattering have virtuality at most of order $m^{2}$.
Then the important integration region for $q_{1}^{+}$ is
$|q_{1}^{+}| \lesssim m^{2}/Q$ so that propagators in the $B$ system
are not far off shell.  Similarly, the important integration region for
$q_{1}^{-}$ is $|q_{1}^{-}| \lesssim m^{2}/Q$.  Then $\left|q_{1}^{+}
q_{1}^{-}\right| \lesssim m^{4}/Q^{2} \ll m^{2} \sim q^{2}_{T}$, so we
can simplify our problem by replacing
\begin{equation}
q_{1\mu }q_{1}^{\mu }= 2 q_{1}^{+}q_{1}^{-} - q_{1T}^{2} \rightarrow -
q^{2}_{1T}
\label{softprops}
\end{equation}
in the gluon propagator.  With this replacement for $q_1$, the soft
gluon line has $q_{1}^{\mu }q_{1\mu }< 0$, so that it can never cross
the final state cut. (This is just to reduce the number of cuts
considered. Our standard analysis \cite{CSS88} does not use this
assumption, which allows us to consider also $|q_{1T}^{2}| \ll m^{2}$.)
Exactly similar remarks apply to the soft gluon, of momentum
$q_2$, in the other ladder.

\begin{figure}
\centerline{\DESepsf(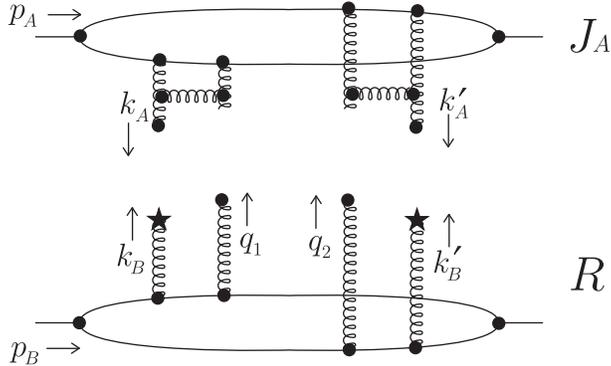 width 8 cm)}
\caption{The basic diagram split into subdiagrams $J_{A}$ and $R$
corresponding to our chosen region of the loop momenta.}
\label{split}
\end{figure}

Now, following the argument in Ref.~\cite{CSS88}, we break up our graph
into $J_{A}$ and $R$, as shown in Fig.~\ref{split}. The jet subgraph
$J_{A}$ contains all of the lines near the mass shell with large plus
momenta, $k^{+} \gtrsim Q$. The remaining subdiagram contains all of the
other lines, including another jet subgraph with lines with large minus
momenta, the two soft lines carrying momenta $q_{1}^{\mu }$ and
$q_{2}^{\mu }$, and the hard subgraph.

\section{Cancellation of the Ladder Region}

We now show the cancellation of the ladder region in the sum over cuts
of Fig.\ \ref{basic}. We proceed in a number of steps that follow the
arguments of Ref.\ \cite{CSS88}.  These start with some approximations
that give power-suppressed errors.

{\it Step 1.} Neglect $k^{+}_{B}$ and $k_B^{\prime +}$ in $J_{A}$ so
that subgraph $R$ includes $\int dk_{B}^{+} dk_{B}^{\prime +}$. 
Similarly, neglect $k_{A}^{-}$ and $k_A^{\prime -}$in $R$ so that
subgraph $J_{A}$ includes $\int dk_{A}^{-} dk_{A}^{\prime -}$.

{\it Step 2.} Neglect $q_{1}^{+}$ and $q_{2}^{+}$ in $J_{A}$. This
gives
$\bar{R} = \int dq_{1}^{+} dq_{2}^{+}\ R \ 
\delta(k_A^{+}+{k^\prime_A}^{+}+q_1^++q_2^+)$.
Also take
$J_{A}^{\mu _{1} \mu _{2}}\ R_{\mu _{1} \mu _{2}} \rightarrow
J_{A}^{++} R_{++}$.

{\it Step 3.} Show that $\bar{R} $ is independent of which side of
the final state cut the upper ends of the two soft gluons are on.

\begin{figure}
\centerline{\DESepsf(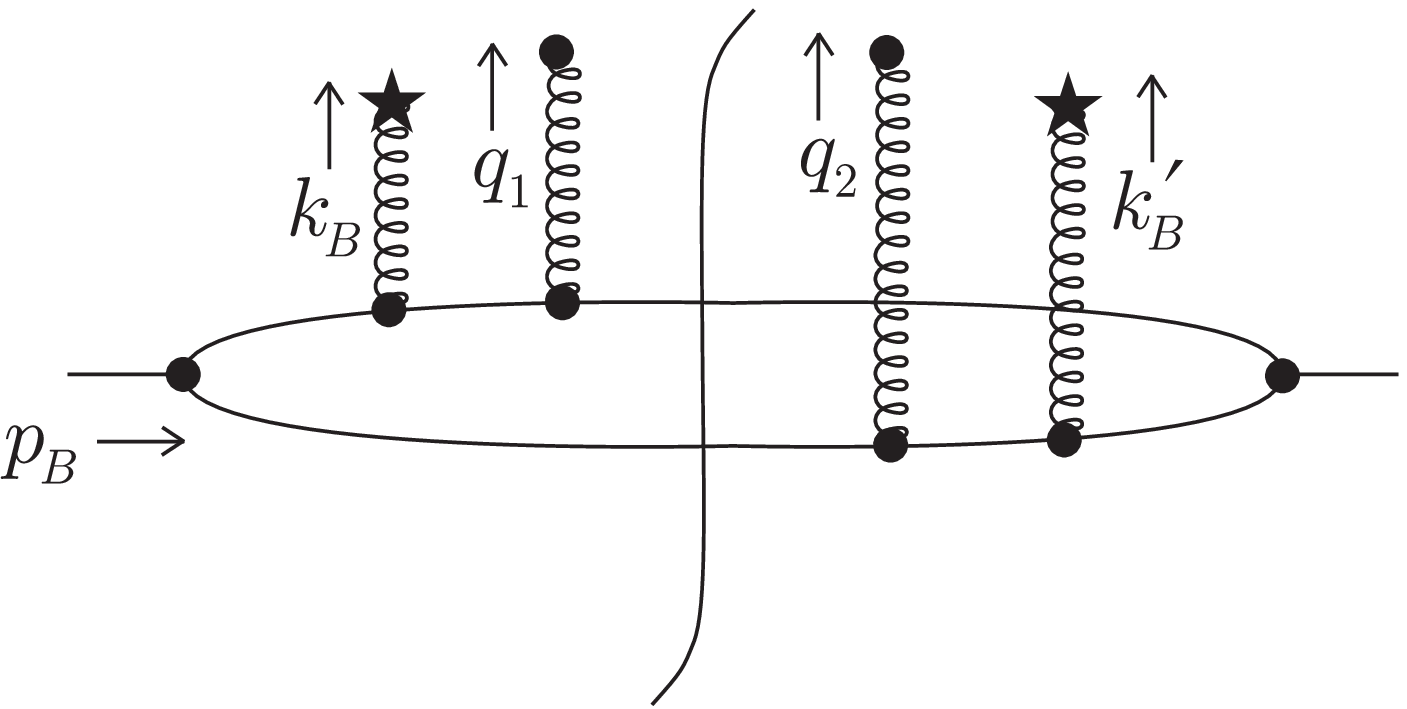 width 7 cm)\
            \DESepsf(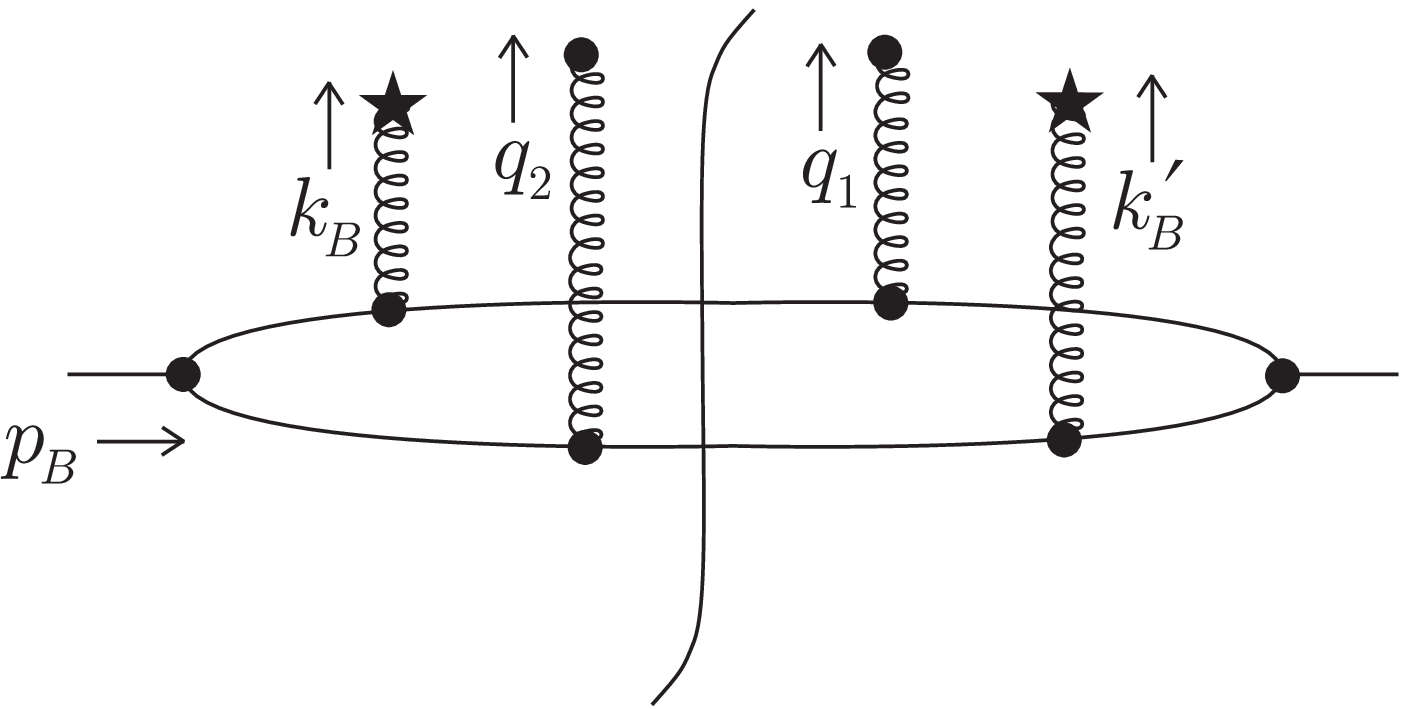 width 7 cm)}
\smallskip
\centerline{\DESepsf(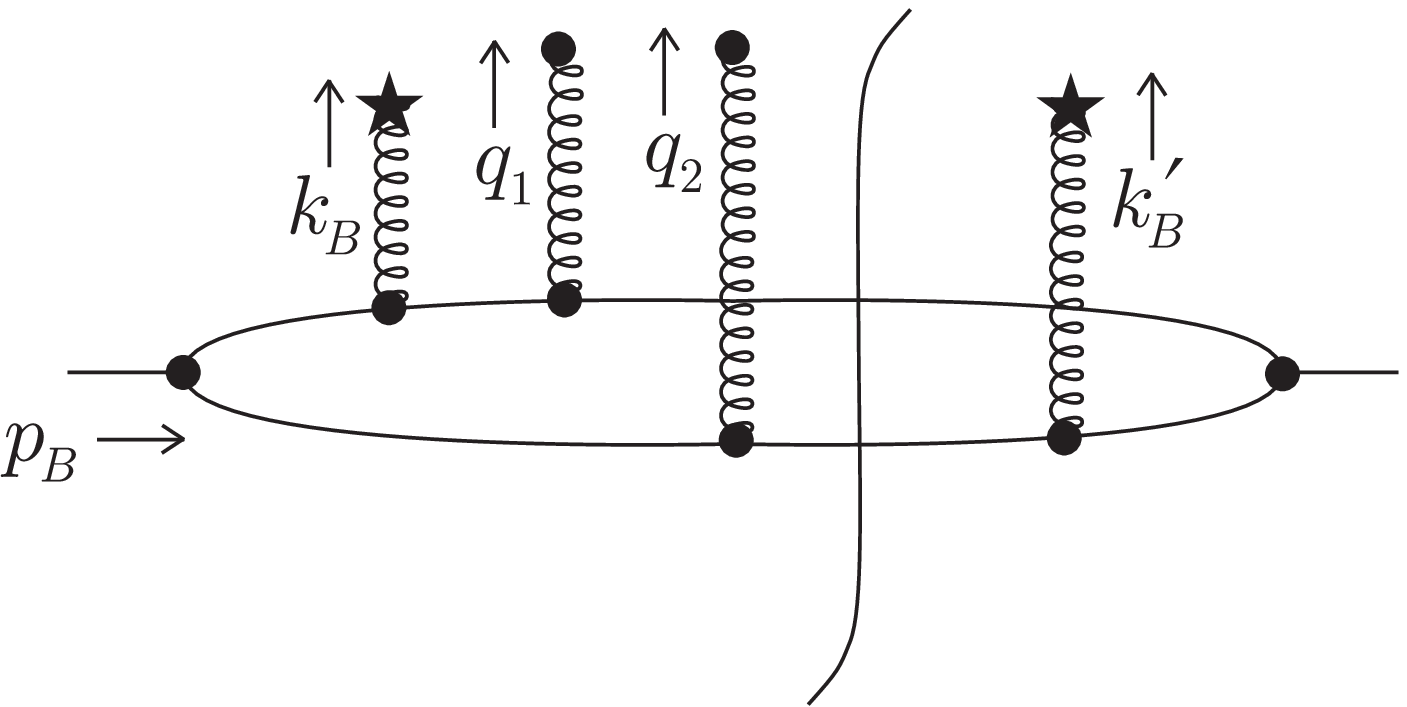 width 7 cm)\
            \DESepsf(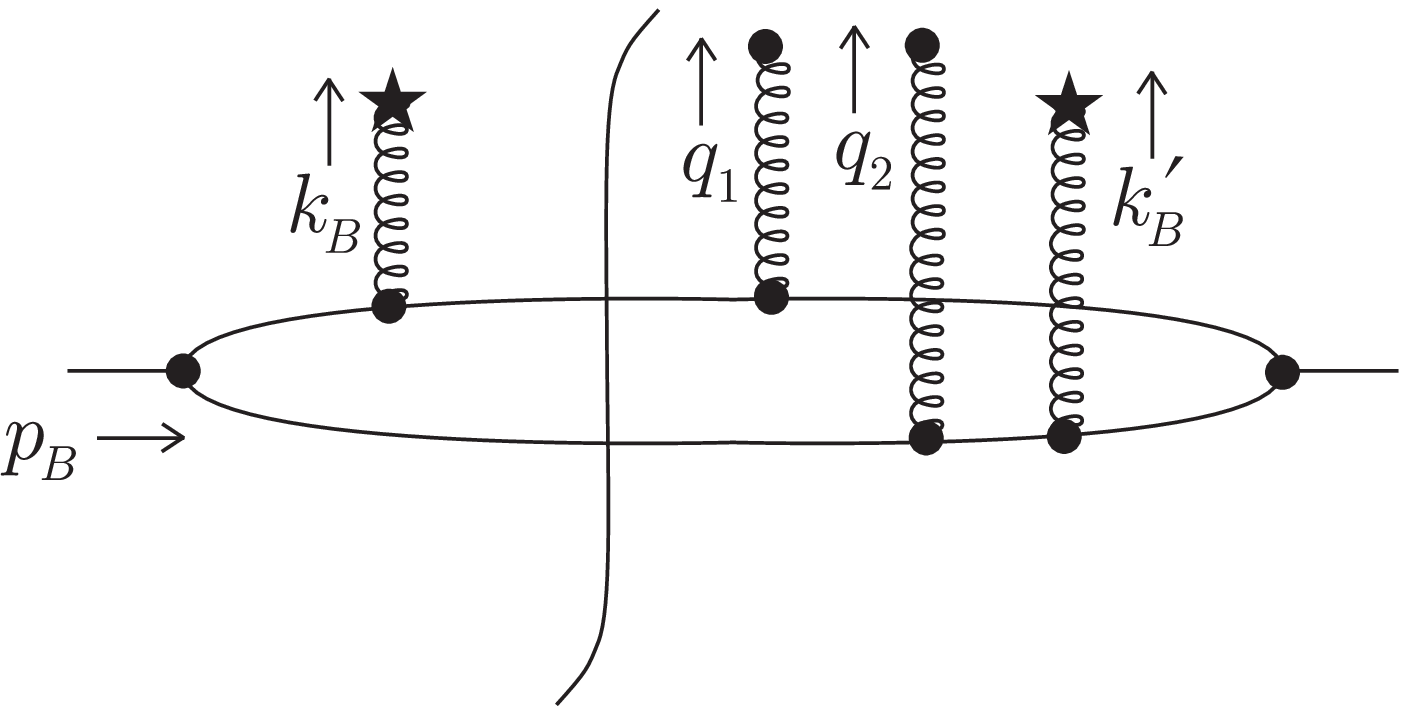 width 7 cm)}
\caption{The four cuts of subgraph $R$.}
\label{rcut}
\end{figure}

Steps 4 and 5 will be explained later and illustrated by explicit
calculation.  Before proceeding to them, we need to discuss {\it Step
3} in detail, since it is not obviously correct. We need to show that
the four cuts of $R$ shown in Fig.~\ref{rcut} give equal contributions
to $R$.

In Ref.~\cite{CSS88}, we show this by writing the diagrams in $x^{-}$
ordered perturbation theory.  Here we have an especially simple
situation, so we can simplify the argument.  Consider, for example,
the first graph in Fig.~\ref{rcut}.

\begin{figure}
\centerline{\DESepsf(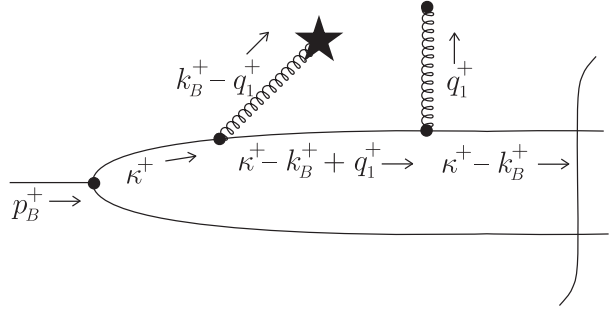 width 8 cm)}
\caption{Routing of plus momentum for first cut of $R$}
\label{rcutrouting}
\end{figure}

We can route $q_{1}^{+}$ and  $q_{2}^{+}$ differently in different graphs
because we are integrating over $k^{+}_{B}$ and we can shift the origin of
the $k^{+}_{B}$ integration. In this particular graph, we choose to route
$q_{1}^{+}$ directly to the left hand hard interaction, as illustrated in
Fig.~\ref{rcutrouting}.

With this momentum routing, we perform the $q_{1}^{+}$ integration
first. After accounting for the replacement (\ref{softprops}), we see
that there are two poles in the complex $q_{1}^{+}$ plane, one from the
quark line carrying momentum $\kappa^+ - k_B^+ + q_{1}^{+}$ and one
from the gluon line carrying momentum $k_B^+ - q_{1}^{+}$. In the
ladder region of integration, the quark lines in the figure all carry
positive momentum fractions $\ell ^{-}/p_{B}^{-}$ from left to right.
Then the two poles in the $q_{1}^{+}$ plane are on opposite sides of
the real $q_{1}^{+}$ axis. We close the $q_{1}^{+}$ integration contour
in the lower half plane to put the quark line on shell. The result is
the same as that obtained by replacing
\begin{equation}
{ i \over k^{2} - m^{2} + i\epsilon }
\to 2\pi \,\delta (k^{2}-m^{2}),
\end{equation}
just as if the quark were crossing a final state cut.

\begin{figure}
\centerline{\DESepsf(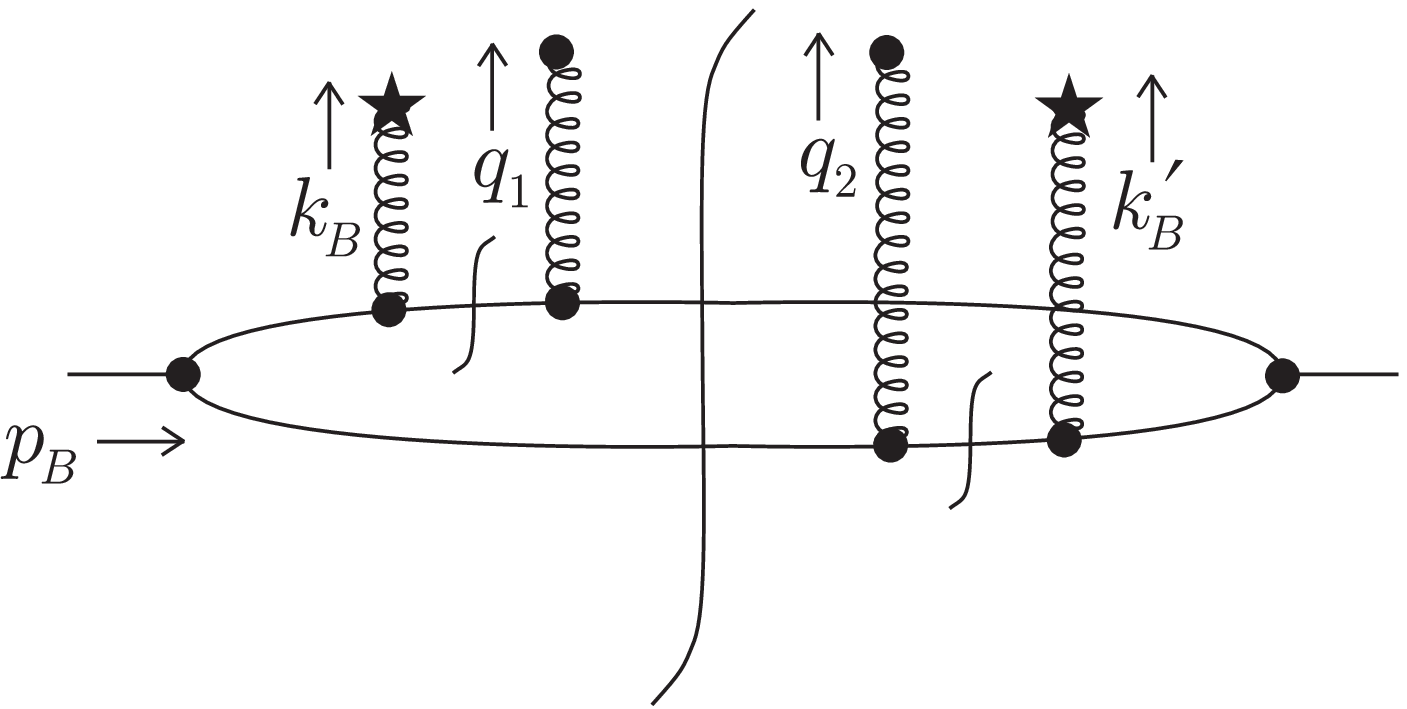 width 7 cm)\
            \DESepsf(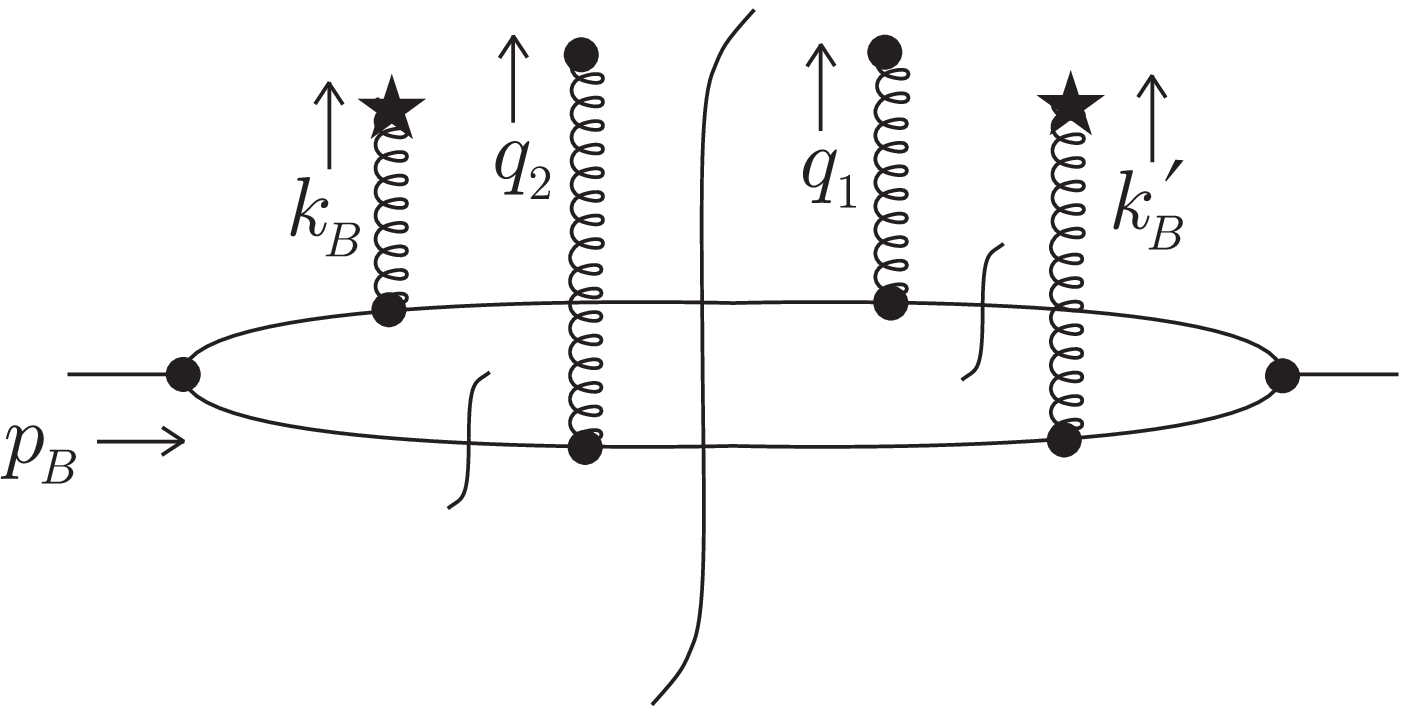 width 7 cm)}
\smallskip
\centerline{\DESepsf(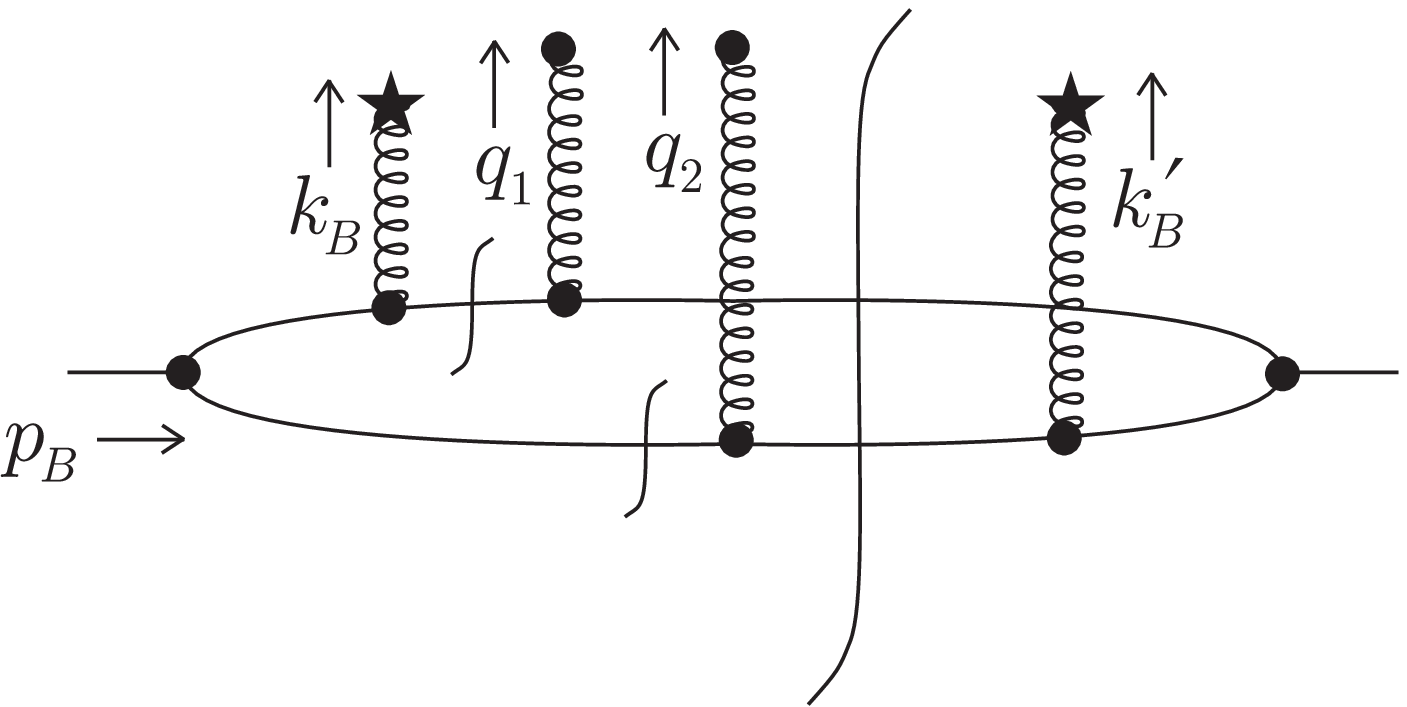 width 7 cm)\
            \DESepsf(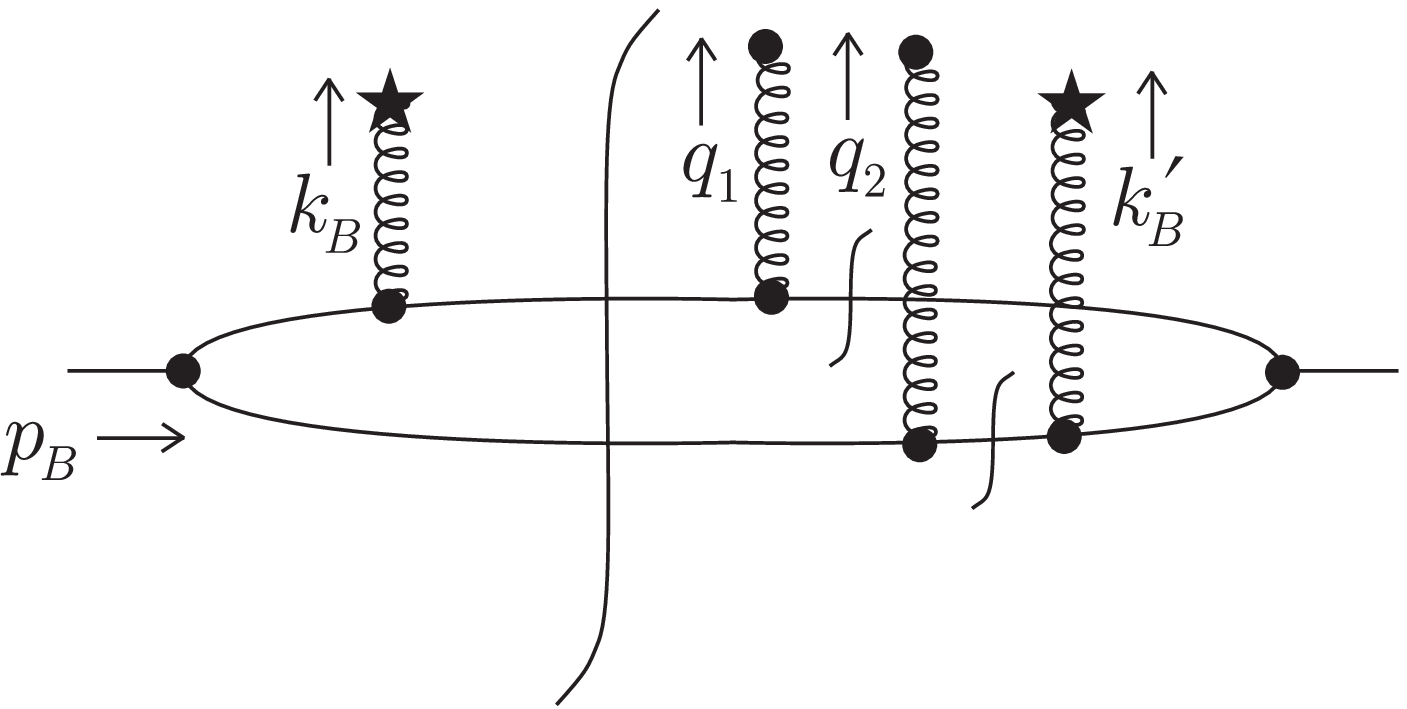 width 7 cm)}
\caption{The four cuts of subgraph $R$ after performing the $q_1^+$
and $q_2^+$ integrations.}
\label{rcutcut}
\end{figure}

We make a similar choice for the $q_{2}^{+}$ integration.
The result is shown in the first graph in Fig.~\ref{rcutcut}, where we
indicate the 
factors $\delta (k^{2}-m^{2})$ with ``cut'' symbols.

Now we do this for each of the four graphs. We get the results shown in
Fig.~\ref{rcutcut}.  Evidently, all four of these cut graphs are equal,
except for the factors of $i$ associated with the vertices and the
uncut propagators.  But to go from one graph to another we transfer one
or more uncut propagators and the same number of vertices from one side
of the final-state cut to the other. Hence the different cuts are in
fact equal.  QED.

\bigskip

{\it Step 4.} Now we look at $J_{A}$. In Ref.~\cite{CSS88} we argue that,
after summing over cuts, we can replace $q_{1}^{T} \to 0$ and $q_{2}^{T} \to 0$
in $J_{A}$. This involves showing (by writing the diagrams in $x^{-}$
ordered perturbation theory) that

\smallskip\noindent a) after summing over cuts, poles in $q_{1}^{-}$ and
$q_{2}^{-}$ near the origin are in the lower half planes;

\smallskip\noindent b) we can deform the $q_{i}^{-}$ integration contours
far into the upper half planes;

\smallskip\noindent c)  on the deformed contours,
if $\ell ^{\mu }$ is the momentum of a propagator through which
$q_{i}^{\mu }$ flows, then $\ell ^{-}$ becomes large:
$\ell ^{+} \ell ^{-} \gg \ell _{T}^{2}$; thus we
can set $q_{i}^{T} \to 0$;

\smallskip\noindent d) finally, we can return the contours to the real
axes.

\bigskip
{\it Step 5.} The argument in Ref.~\cite{CSS88} proceeds to combine the
results from different Feynman graphs, using gauge invariance, to
finally write the result in a factored form.

In the present case we have an especially simple situation, so we can
simplify the argument. In fact, the contribution from the ladder region
to the graph we are considering vanishes at leading order in $1/Q$ by
itself. Thus we do not need {\it Step 5}. We will therefore make a
straightforward analysis of the graphs to show that

\medskip {\it Step 4${\,}^{\prime}$}. After summing over cuts, $J_{A} =
{\cal O}(1/Q)$ in the ladder region.  

\smallskip
Since subgraph $R$ is the same no matter where we draw the cut, this
implies that the sum over cuts of the original graph vanishes at the
leading power of $1/Q$.

\begin{figure}
\centerline{\DESepsf(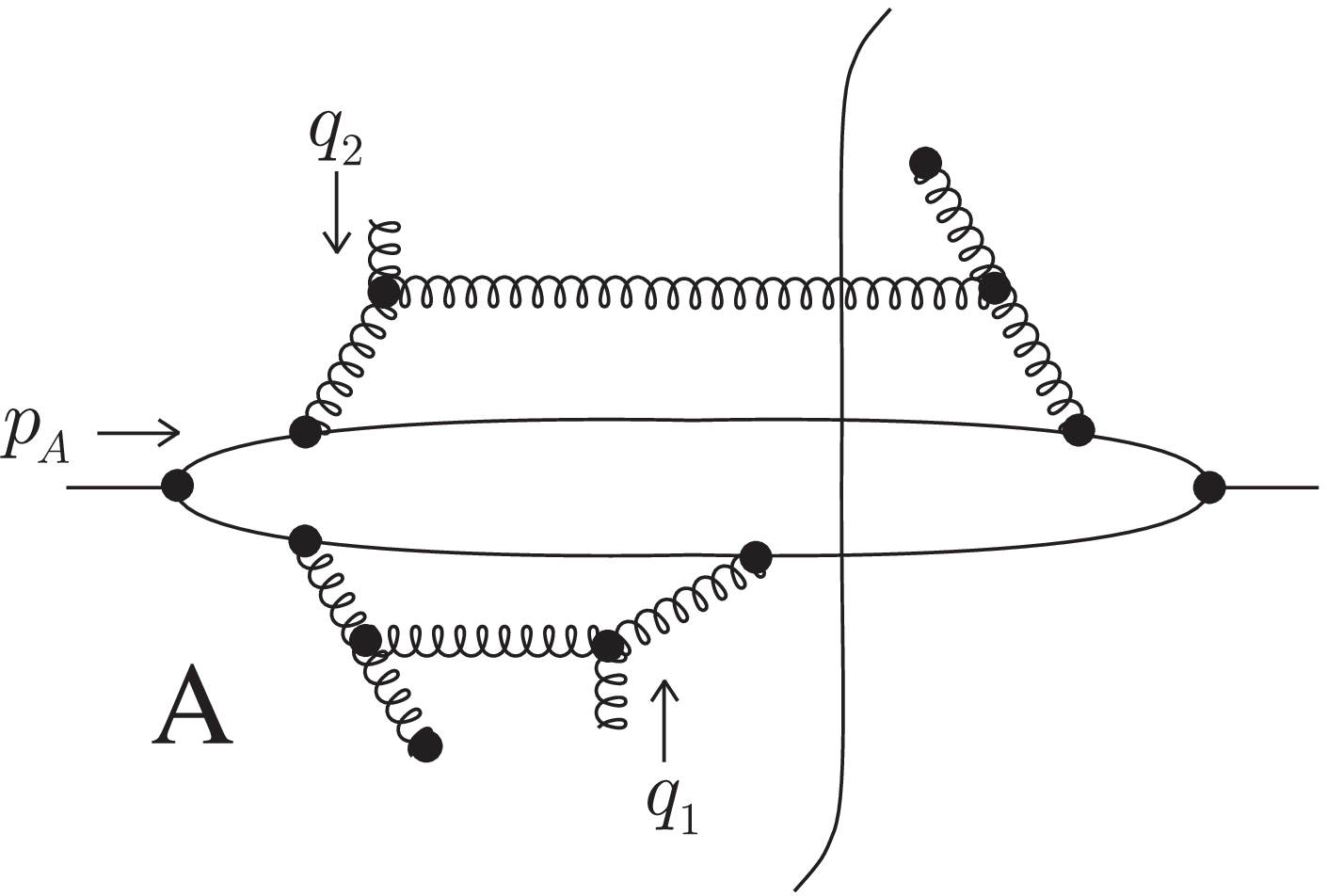 width 7 cm)\
            \DESepsf(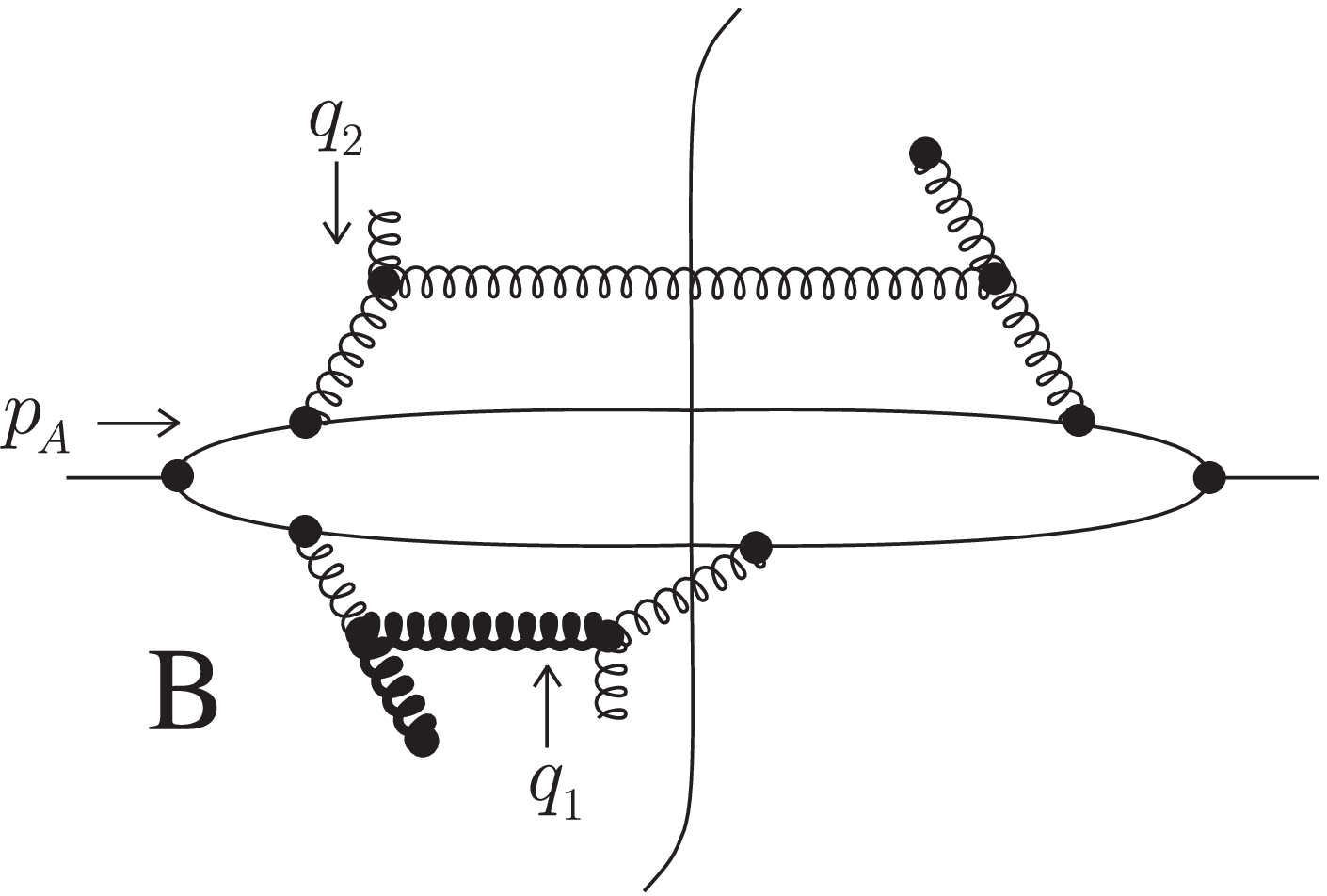 width 7 cm)}
\smallskip
\centerline{\DESepsf(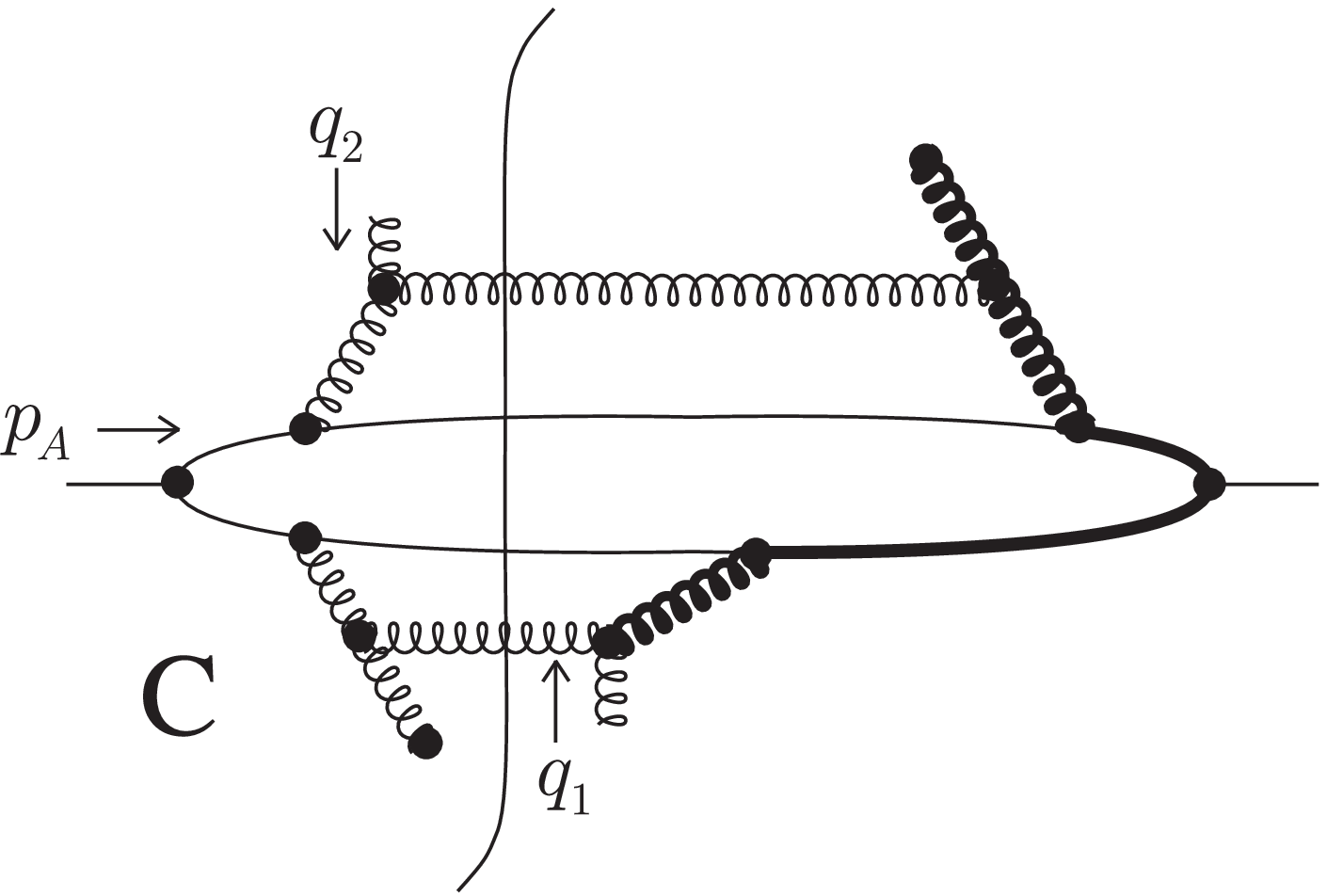 width 7 cm)\ }
\caption{Three cuts of subgraph $J_{A}$.}
\label{jetcut}
\end{figure}

We now show explicitly that the sum over cuts of the jet subgraph
$J_{A}$ gives zero. There are nine cuts of $J_{A}$ to consider. We look
first at the three cuts shown in Fig.~\ref{jetcut}. Call these cut
graphs $A$, $B$, and $C$. We have oriented the lines to indicate the
direction of flow of the $+$ component of momentum; recall that the $+$
components of $q_{1}$ and $q_{2}$ have been set to zero in $J_{A}$.

\begin{figure}
\centerline{\DESepsf(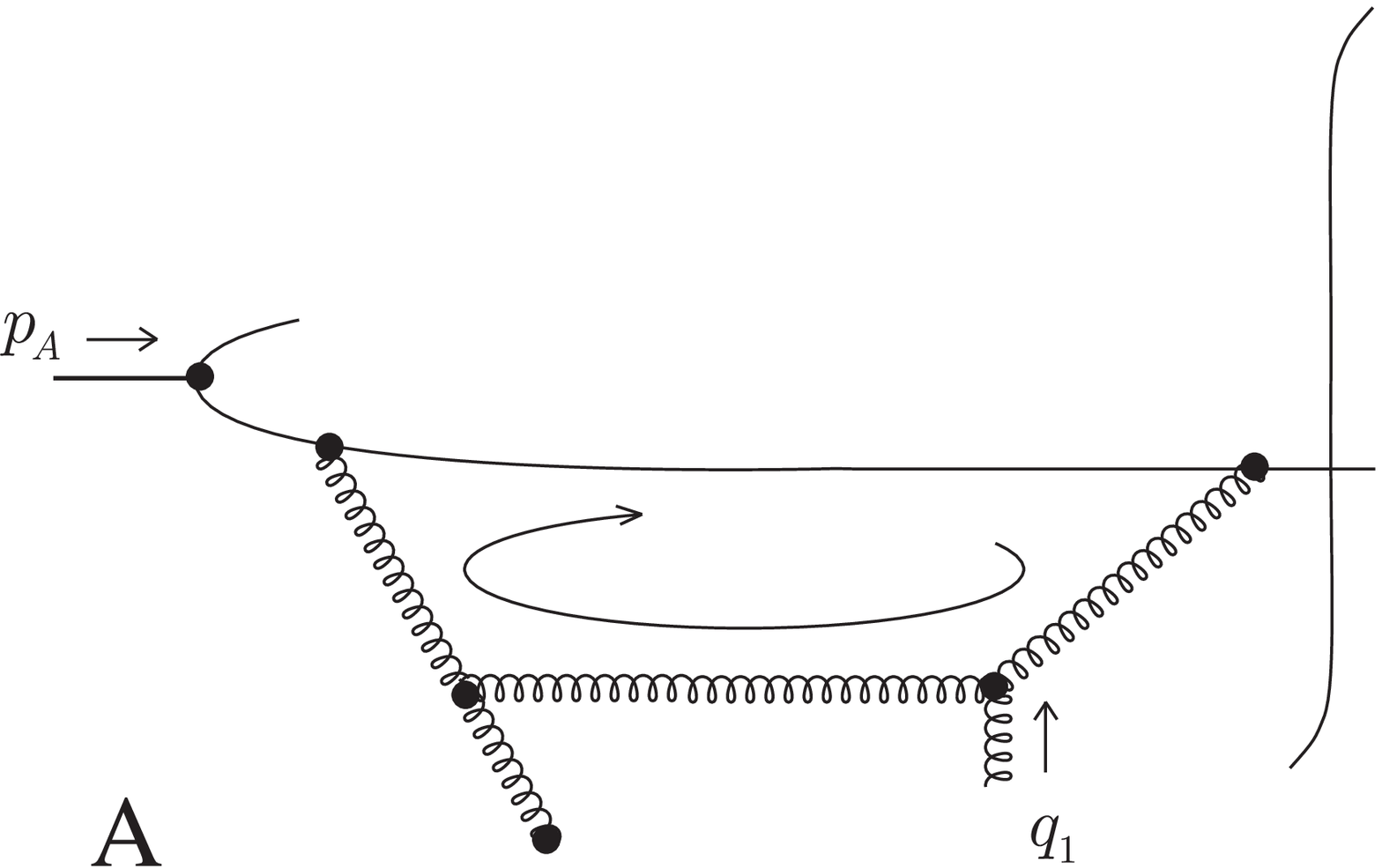 width 7 cm)\ \ \ \
            \DESepsf(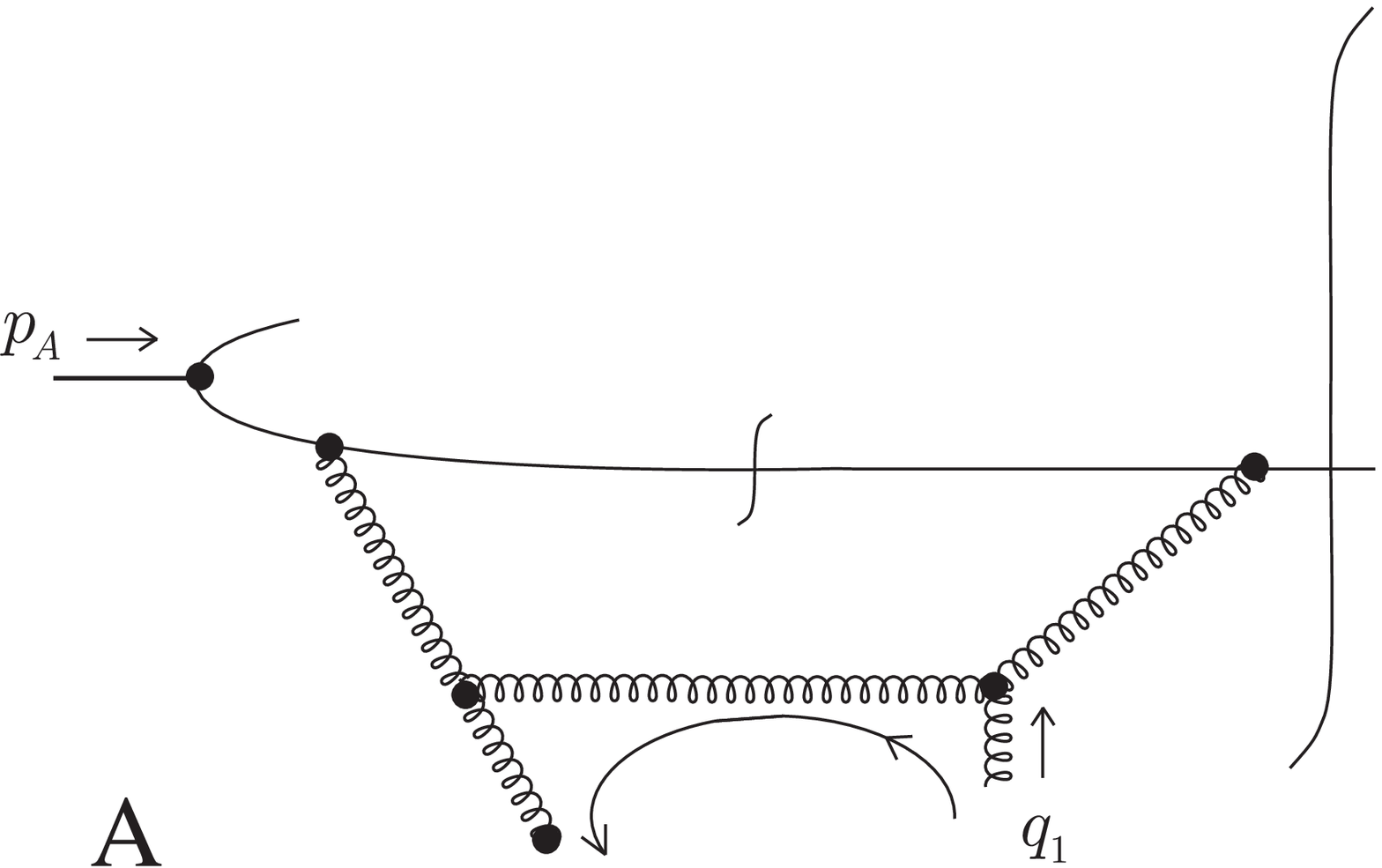 width 7 cm)}
\caption{Treatment of cut A of $J_A$.}
\label{jetcutafate}
\end{figure}

\begin{figure}
\centerline{\DESepsf(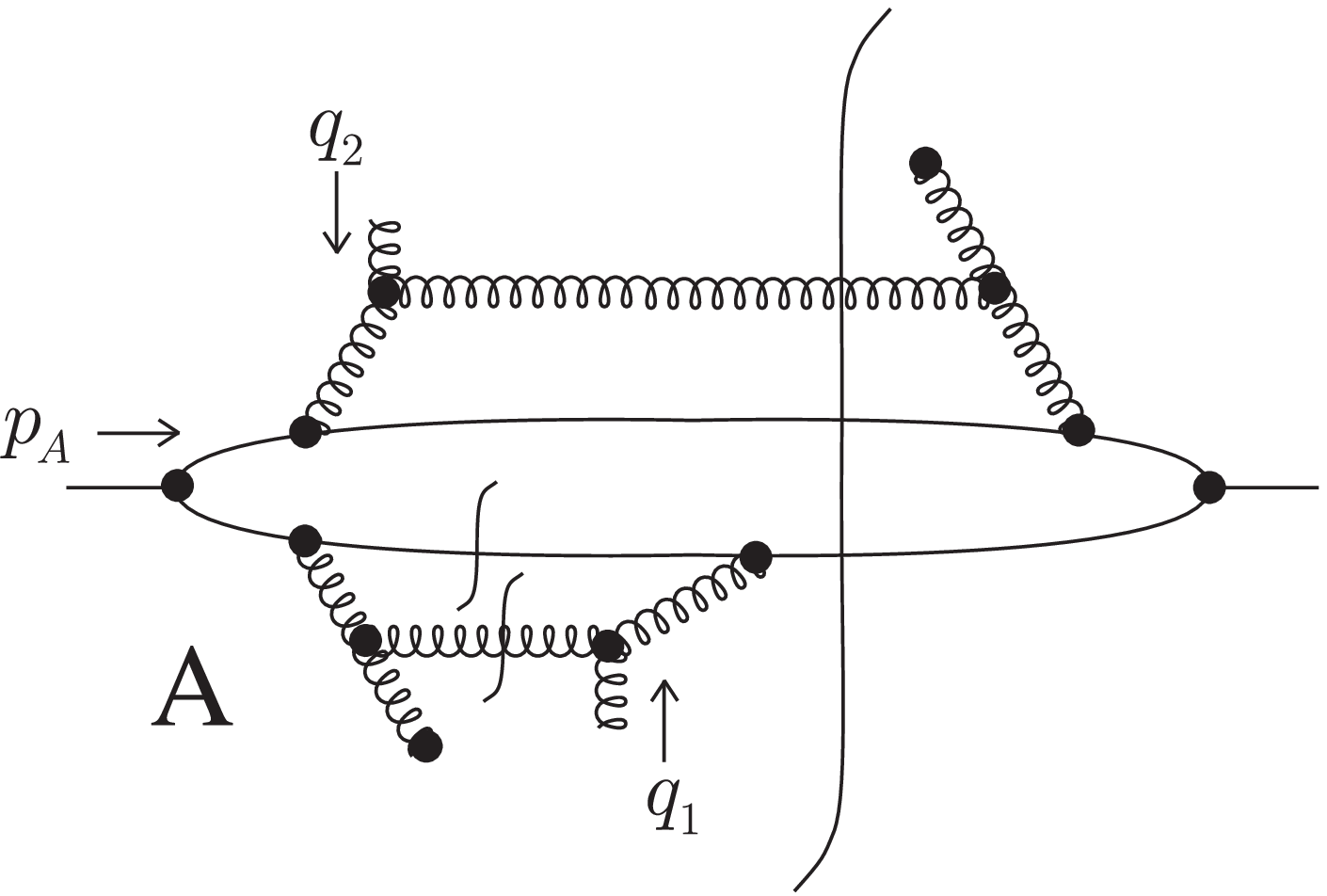 width 7 cm)\
            \DESepsf(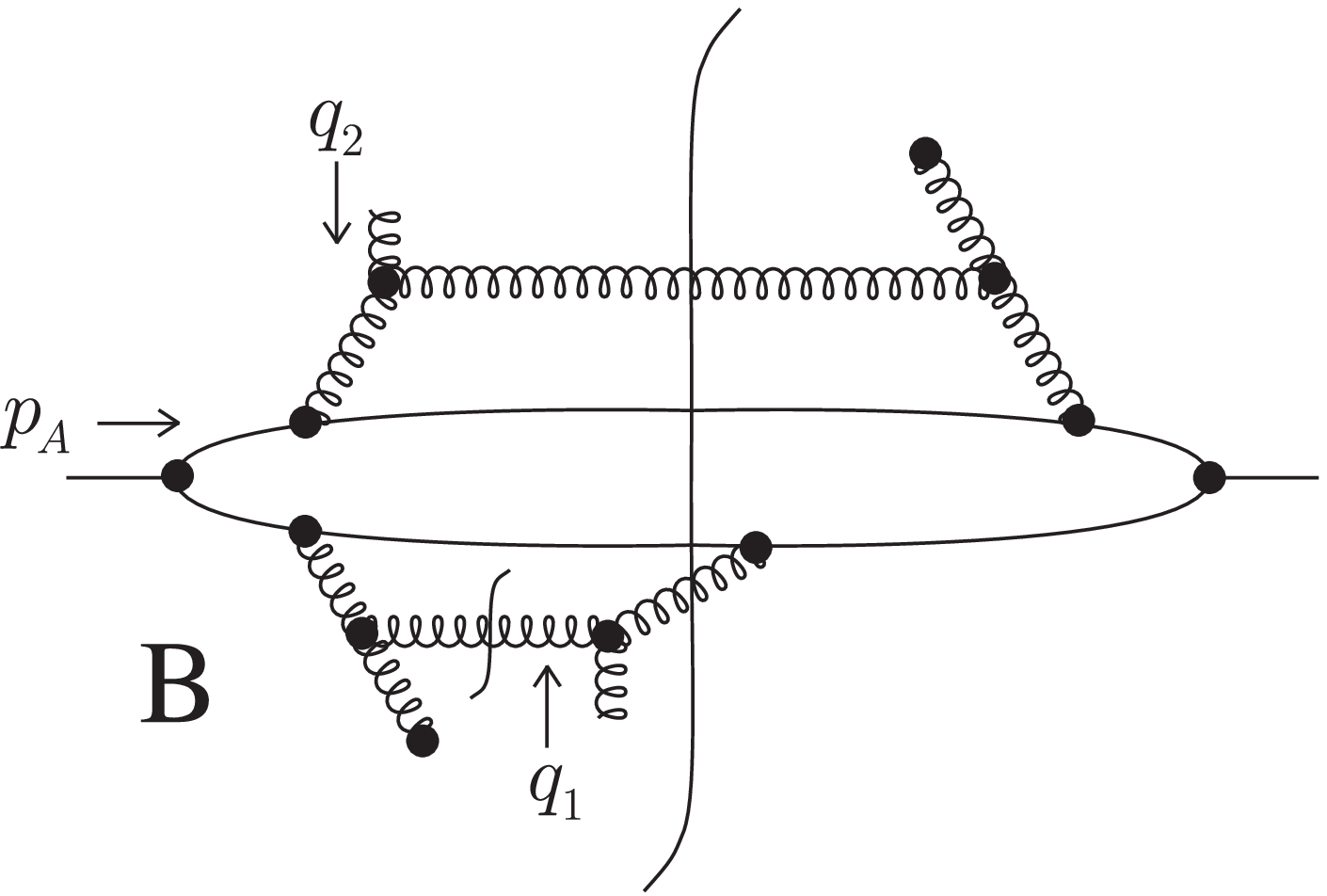 width 7 cm)}
\smallskip
\centerline{\DESepsf(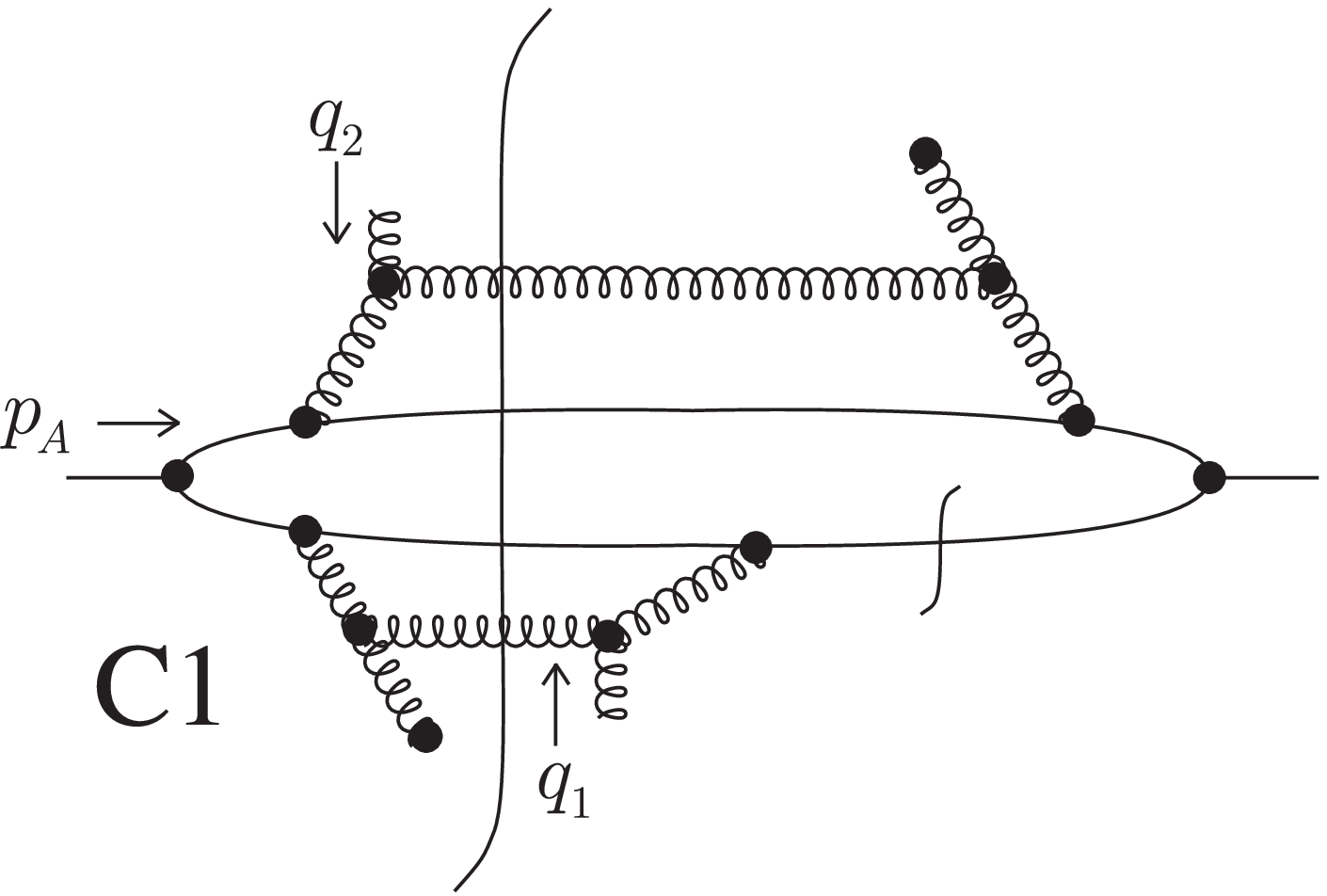 width 7 cm)\
            \DESepsf(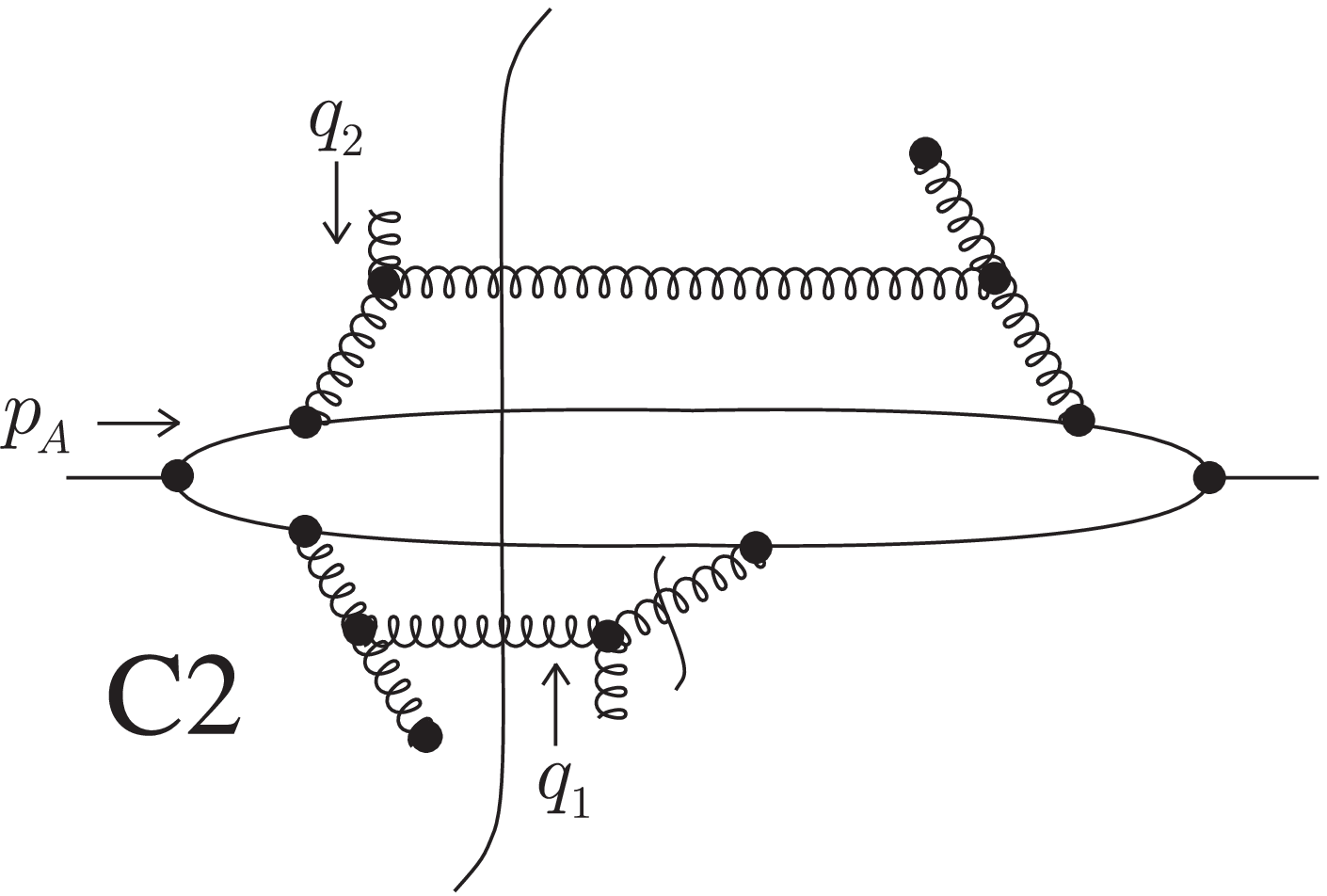 width 7 cm)}
\caption{Three cuts of subgraph $J_{A}$ after performing minus
integrations.}
\label{jetcutcut}
\end{figure}

We consider first cut graph $A$. This cut graph has a virtual loop,
shown in more detail in Fig.~\ref{jetcutafate}. Let loop momentum
$\ell^-$ circulate around this loop as shown in the first part of
Fig.~\ref{jetcutafate}. Since, in the ladder integration region,
positive $+$ momentum flows from left to right through the
propagators in the figure, there are three poles in the upper half
$\ell^-$ plane and one in the lower half $\ell^-$ plane. We close the
contour in the lower half $\ell^-$ plane. This amounts to putting the
quark line on shell, as indicated by the cut symbol in second part of
Fig.~\ref{jetcutafate}. Next, we route $q_{1}^{-}$ to the hard vertex
through the two gluon propagators, as depicted in the second part of
Fig.~\ref{jetcutafate}. There are two poles in the $q_{1}^{-}$ plane.
We perform the $q_{1}^{-}$ integral by closing the contour in the
upper half plane, which amounts to putting the first gluon through
which $q_{1}^{-}$ flows on shell. The result is indicated in the first
graph, $A$, of Fig.~\ref{jetcutcut}.

In cut graph $B$, we route $q_{1}^{-}$ to the hard vertex through the
two gluon propagators, indicated by heavy lines in Fig.~\ref{jetcut}.
Then closing the $q_{1}^{-}$ contour in the upper half plane gives the
result indicated in the second graph, $B$, of Fig.~\ref{jetcutcut}

In cut graph $C$, we route $q_{1}^{-}$ to the hard vertex to the right,
through a gluon propagator, two quark propagators, and two more gluon
propagators, indicated with heavy lines. There are five poles
in the $q_{1}^{-}$ plane, two in the upper half plane and three in the
lower half plane. We perform the $q_{1}^{-}$ integral by closing the
contour in the upper half plane. Then there are two terms. The two terms
are indicated in the third and fourth graphs, $C1$ and $C2$, of
Fig.~\ref{jetcutcut}.

Now we note that $C1$ cancels $A$, while $C2$ cancels $B$: in each of
the canceling pairs of graphs, the propagators and vertices are the
same up to sign and there are an odd number of sign changes resulting
from moving uncut propagators and vertices from one side of the final
state cut to the other.

There are more cut graphs. For each way that we cut the right hand
ladder, there are three ways to cut the left hand ladder. Thus we
could label the graphs $G(n,m)$ with $n = A,B,C$ denoting the cut for
the left hand ladder and $m = A,B,C$ denoting the cut for the right
hand ladder. In this notation, what we have seen is that $G(A,C) +
G(B,C) + G(C,C) = {\cal O}(1/Q)$. The same argument shows that
\begin{equation}
\sum _{n=A, B, C} G(n,m) = {\cal O}(1/Q) \quad m = A, B, C.
\end{equation}
A similar argument also shows that
\begin{eqnarray}
\sum _{m=A, B, C} G(n,m) = {\cal O}(1/Q) \quad n = A, B, C.
\end{eqnarray}
Thus there is a double cancellation.

We conclude that the sum over all cuts of Fig.\ \ref{basic} is ${\cal
O}(1/Q)$ and we do not have a violation of factorization.

\section{Extensions of the argument}

Gotsman {\it et al.} consider more elaborate diagrams than the
diagram of Fig.\ \ref{basic} that we have analyzed here. For instance, they
consider exchanges of gluon ladders with arbitrary numbers of rungs. In
addition, they concentrate on the region of integration space in which
the rapidities of the rungs are strongly ordered. It should be evident
that the calculation given above could be applied to more complicated
diagrams. There would be little to change in the
pattern of cancellation found here,
which illustrates the general argument of Ref.\ \cite{CSS85,CSS88},
although the
notation would have to be more complicated. We draw attention to the
fact that although factorization applies in the strongly ordered region
that gives the leading small $x$ logarithms, factorization also applies
in other regions: strong ordering of rapidities is not essential to the
argument.

The argument presented here applies when the exchanged gluons have
transverse momentum of order $m$, and when none of the lines
outside of the hard scattering has virtuality much larger than
$m^{2}$.  This is also the region discussed in
Ref.~\cite{levinetal}.  Other regions than the one we considered
here are correctly treated by other parts of the factorization
proof in \cite{CSS85,CSS88}.

Let us close with a few additional comments relating our work to that of Ref.\
\cite{levinetal}. Gotsman {\it et al.} use modified AGK cutting rules
to perform their calculation, finding a remainder after summing over
final states. Such an analysis, however, assumes that the incoming
particles to which the ladders attach are on-shell.  In addition,
we believe that this analysis is exact only at the level of leading
logarithms.  However, their noncancelling result (Eq.\ (8) in Ref.\
\cite{levinetal}), which is proportional to the square of the real
part of their single-ladder exchange amplitude, is suppressed by two
logarithms compared to leading logarithms.

We may note two specific differences between our analysis and that of
Gotsman {\it et al.}, both of which are important beyond the leading
logarithmic approximation.  First, they ignore certain final-state cuts,
for example cut B in Fig.\ \ref{jetcut}. This cut crosses the side of a
ladder. Such a cut is not allowed in the leading logarithm approximation,
but it is present when one goes beyond leading logarithms. Observe that
the cut in question would not be kinematically allowed if the gluon ladder
coupled to on-shell quark lines instead of internal off-shell lines.
Second, Gotsman {\it et al.} appear to have ignored the loop integral
that we label $k_B^+$,  which we use to prove that the subgraph $R$ is
independent of the final state cut of $J_A$ ({\it Step 3}).  Both the
extra cuts and the integral over $k_B^+$ are necessary to evaluate
the ladder region fully.  The proof of factorization that we gave in
Ref.~\cite{CSS85,CSS88} applies to the full leading-power contribution
and not just to leading logarithms.  It is this more general method that
we have explained in the context of particular Feynman graphs.

Although we believe that factorization works, we should emphasize our
concern that improvements are needed in the rigor of existing proofs of
factorization \cite{CSS88,CSSrv}. Conspicuously lacking are iterative
algorithms for demonstrating (\ref{factth}) along the lines of those
available for proofs of renormalization.  Our presentation here is not
a step in this direction, rather it is an illustration of
already-existing arguments.  We believe, however, that it is possible
to construct such an algorithm ({\it Cf.} Ref.\ \cite{CSTkachov}).
Improvements in the proof would help both to increase confidence in
factorization and to clarify its variant applications, for instance to
transverse momentum distributions \cite{qtfact} in the range of
$Q_{T}\ll M$.  We therefore welcome the motivation provided by 
\cite{levinetal} to reopen the discussion of
these important issues.

\section*{Acknowledgments}

We thank the Aspen Center for Physics for its hospitality during the
formative stages of this analysis. This work was supported by the U.\
S.\ Department of Energy grants DE-FG02-90ER-40577 and
DE-FG03-96ER40969 and the U.\ S.\ National Science Foundation, grant
PHY9722101.

\end{document}